\documentclass[journal]{IEEEtran}
\usepackage{fmtcount}
\usepackage{psfrag}
\usepackage{subfigure}
\usepackage{caption}
\usepackage{amssymb}
\usepackage{amsmath}
\usepackage{caption}
\usepackage{graphicx}
\usepackage[usenames,dvipsnames]{xcolor}

\begin{document}

\graphicspath{{FIGURES/}}

\title{Analog Signal Processing}

\author{Christophe~Caloz,~\IEEEmembership{Fellow,~IEEE,}
        Shulabh~Gupta,~\IEEEmembership{Member,~IEEE,}
        Qingfeng~Zhang,~\IEEEmembership{Member,~IEEE,}
        and~Babak~Nikfal,~\IEEEmembership{Student~Member,~IEEE}}

\maketitle

\begin{abstract}
Analog signal processing~(ASP) is presented as a systematic approach to address future challenges in high speed and high frequency microwave applications. The general concept of ASP is explained with the help of examples emphasizing basic ASP effects, such as time spreading and compression, chirping and frequency discrimination. Phasers, which represent the core of ASP systems, are explained to be elements exhibiting a frequency-dependent group delay response, and hence a nonlinear phase response versus frequency, and various phaser technologies are discussed and compared. Real-time Fourier transformation (RTFT) is derived as one of the most fundamental ASP operations. Upon this basis, the specifications of a phaser -- resolution, absolute bandwidth and magnitude balance -- are established, and techniques are proposed to enhance phasers for higher ASP performance. Novel closed-form synthesis techniques, applicable to all-pass transmission-type cascaded C-section phasers, all-pass reflection-type coupled resonator phasers and band-pass cross-coupled resonator phasers are described. Several applications using these phasers are presented, including a tunable pulse delay system, a spectrum sniffer and a real-time spectrum analyzer~(RTSA). Finally, future challenges and opportunities are discussed.
\end{abstract}
%
%
\IEEEpeerreviewmaketitle

\section{Introduction and Motivation}\label{sec:intro_motiv}

Today's exploding demand for faster, more reliable and ubiquitous radio systems in communication, instrumentation, radar and sensors poses unprecedented challenges in microwave and millimeter-wave engineering.

Recently, the predominant trend has been to place an increasing emphasis on digital signal processing (DSP). However, while offering device compactness and processing flexibility, DSP suffers of fundamental drawbacks, such as high-cost analog-digital conversion, high power consumption and poor performance at high frequencies.

To overcome these drawbacks, and hence address the aforementioned challenges, one might possibly get inspiration from ultrafast optics~\cite{Saleh_Teich_FP}. In this area, ultra-short and thus huge-bandwidth electromagnetic pulses are efficiently \emph{processed in real time} using \emph{analog} and \emph{dispersive} materials and components~\cite{Azana_JQE_2000,Berger_EL_2000,Yao_PTL_2000}.

We speculate here that the same approach could be potentially applied to high-frequency and high-bandwidth microwave signals, leading to systematic \emph{microwave analog signal processing (ASP)} as an alternative to DSP-based processing, with particular promise for millimeter and terahertz frequency applications.

The paper first explains more specifically what is meant by microwave ASP, covering the basic chirping and frequency discrimination effects, emphasizing the concept of group delay engineering, and presenting the fundamental concept of real-time Fourier transformation. Next, it addresses the  topic of the ``phaser'', which is the core of an ASP system; it reviews phaser technologies, explains the characteristics of the most promising microwave phasers, and proposes corresponding enhancement techniques for higher ASP performance. Based on ASP requirements, found to be phaser resolution, absolute bandwidth and magnitude balance, it then describes novel synthesis techniques for the design of all-pass transmission and reflection phasers and band-pass cross-coupled phasers. Finally, it presents some applications using these phasers, including a tunable pulse delay system for pulse position modulation, a real-time spectrum sniffer for cognitive radio and a real-time spectrogram analyzer for the characterization and processing of nonstationary signals.

\section{What is Microwave Analog Signal Processing?}\label{sec:what_is_ASP}

Microwave ASP might be defined as the manipulation of signals in their pristine \emph{analog} form and in \emph{real time} to realize specific operations enabling \emph{microwave or millimeter-wave and terahertz applications}.

\begin{figure}[h!]
\centering
\includegraphics[width=0.5\textwidth,page=1]{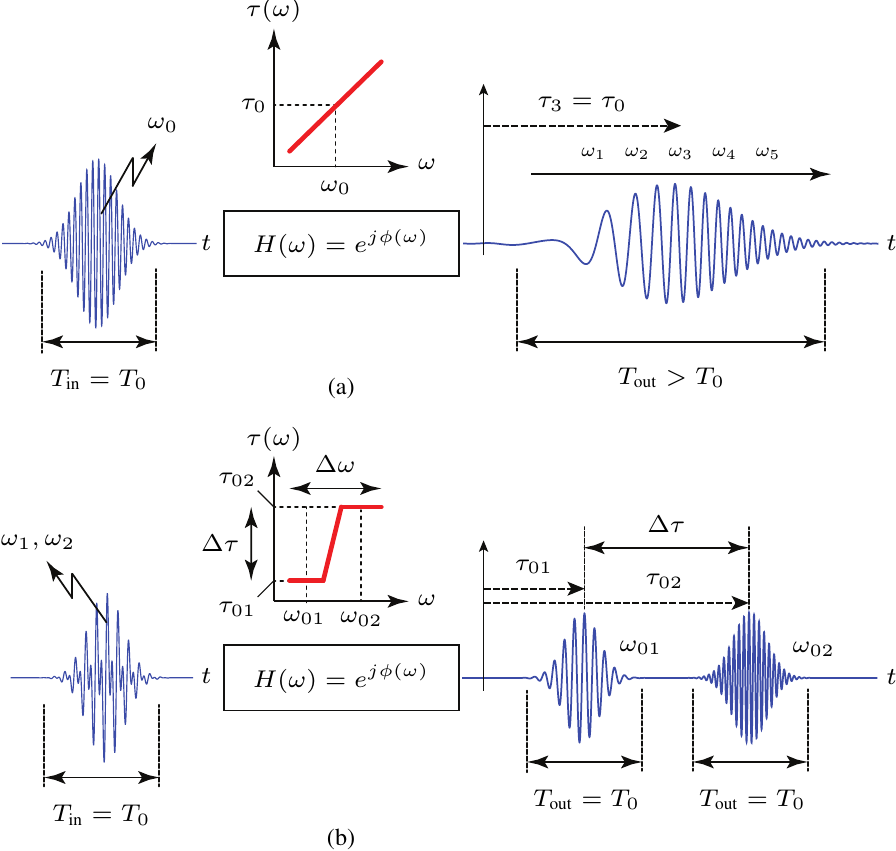}
\caption{Basic effects in ASP. (a)~Chirping with time spreading. (b)~Frequency discrimination in the time domain.}
\label{fig:ASP_basic_effects}
\end{figure}

The essence of ASP might be best approached by considering the two basic effects described in Fig.~\ref{fig:ASP_basic_effects}, chirping with time spreading and frequency discrimination in the time domain. Both effects involve a linear element with transfer function $H(\omega)=e^{j\phi(\omega)}$, which is assumed to be of unity magnitude and whose phase, $\phi(\omega)$, is a nonlinear function of frequency, or whose group delay, $\tau(\omega)=-\partial\phi(\omega)/\partial\omega$, assuming the time-harmonic phasor dependence $e^{j\omega t}$, is a function of frequency. Note that there is no incompatibility between the linearity of the element and the nonlinearity of its phase: the former refers to the independence of the element's response to the \emph{magnitude of the input signal}, whereas the latter refers to the nonlinearity of the element's \emph{phase versus frequency}, which is an \emph{inherent property of the element}, independent of the input signal. Such an element, with \emph{frequency-dependent group delay}, is called \emph{temporally dispersive}, or simply \emph{dispersive} when the context is unambiguous. It is to be noted that \emph{temporal dispersion} contrasts with \emph{spatial dispersion}, occurring for instance in the phenomena of radiation through an aperture or propagation across a periodic structure, where different spatial frequencies, corresponding to scattering in different directions, are generated. The spectrum of $H(\omega)$ is assumed to cover the entire spectrum of the input signal.

In the chirping case, depicted in Fig.~\ref{fig:ASP_basic_effects}(a), a pulse (typically be a gaussian pulse, whose spectrum is also a gaussian function~\cite{Papoulis_FIA}) modulated at an angular frequency~$\omega_0$ is passed through the element $H(\omega)$, which is assumed here to exhibit a positive linear group delay slope over a frequency band centered at the frequency $\omega_0$, corresponding to a group delay $\tau_0$. Due to the dispersive nature of this element, the different spectral components of the pulse experience different delays and therefore emerge at different times. Here, the lower-frequency components are less delayed and therefore emerge earlier than the higher-frequency components, while the center-frequency component appears at the time $\tau_3=\tau_0$. This results in an output pulse whose instantaneous frequency is progressively increasing (would be decreasing in the case of a phaser with a negative group delay slope), a phenomenon called ``chirping'', and which has experienced time spreading~($T_\text{out}>T_0$), accompanied with reduced amplitude due to energy conservation. When undesired, this time spreading effect can be compensated by either using a stepped group delay phaser, as will be shown in the next paragraph, or by \emph{compression}, using phasers of opposite chirps, as will be shown in the tunable pulse delay line application of Sec.~\ref{sec:ASP_apps}. It should be noted that in many applications, time spreading is actually useful. For instance, it belongs to the essence of real-time Fourier transformation, to be presented in Sec.~\ref{sec:RTFT}. As another example, time spreading may be exploited to increase the sampling rate of a signal at fixed clock~\cite{Azana_TMTT_2007s},\cite{Xiang_TMTT_11_2012}.

In the latter case, depicted in Fig.~\ref{fig:ASP_basic_effects}(b), the input pulse is modulated by a two-tone signal, with frequencies $\omega_{01}$ and $\omega_{02}$, and passed through a dispersive element $H(\omega)$ exhibiting a positive stepped group delay, with two steps, centered at $\omega_{01}$ and $\omega_{02}$, respectively. Based on this dispersive characteristic, the part of the pulse modulated at the lower frequency, $\omega_{01}$, is delayed less than the part modulated at the higher frequency, $\omega_{02}$, and hence emerges earlier in time. As a result, the two pulses are \emph{resolved} (or separated) in the time domain, and their respective modulation frequencies may be deduced from their respective group delays from the dispersive law,~$H(\omega)$. Note that with the flat-step law considered here the pulses are \emph{not} time-spread ($T_\text{out}=T_0$), assuming that the pulse bandwidth fits in the flat bands of the steps, since all the spectral components within each band are delayed by the same amount of time.

Reference~\cite{Zhang_APM_TBP} provides a fundamental explanation of group delay dispersion in terms of wave interference mechanisms.

\section{Core of an ASP System: the Phaser}\label{sec:core_ASP_phaser}

The dispersive element $H(\omega)$ in Fig.~\ref{fig:ASP_basic_effects}, which manipulates the spectral components of the input signal in the time domain, is the \emph{core of an ASP system}. We suggest to call such a dispersive element, in which a nonlinear phase versus frequency response is necessary for ASP, a \emph{phaser}. Several terms have been used by the authors and discussed with colleagues from various academic and industrial backgrounds. After much reflection, ``phaser'' imposed itself as the most appropriate term. Table~\ref{tab:phaser_term} lists the four main terms that were considered with their pros and cons, and shows the superiority of ``phaser'' over the other three terms. A phaser ``phases'' as a filter filters: it modifies the phase of the input signal following phase -- or more fundamentally group delay -- specifications, with some magnitude considerations, just in the same way as a filter modifies the magnitude of the input signal following magnitude specifications, with some phase considerations.

It has come to the attention of the authors that a few people seem to dislike the terminology ``phaser.'' One of them asserted that ``this term should be reserved to Star Trek''~[sic]. Well, the fact that ``phaser'' refers to a directed-energy weapon in this popular science fiction series~\cite{Star_Trek} is not considered a major concern by the authors, who judge that confusion between ASP phasers and Star Trek phasers is fairly unlikely!

\begin{table}[h!]
\caption{Comparison of the main terms considered to designate an element of the type $H(\omega)$~in Fig.~\ref{fig:ASP_basic_effects}, following frequency-dependent group delay specifications for ASP.}
\begin{tabular}{c|cc}
  & Pros & Cons \\
  \hline
  \begin{minipage}{1.1cm}
  Phaser
  \end{minipage}
  & \hspace{-0.4cm}\begin{minipage}{4cm}
  \begin{itemize}\vspace{1mm}
    \item natural and unambiguous
    \item counterpart in acoustics~\cite{Smith_W3KP_2010}
    \item general (transfer function \\ - phasing, material, structure,
        device, network, circuit, \\ component, system)
    \item short and elegant
  \end{itemize}\vspace{1mm}
  \end{minipage}
    & \hspace{-0.4cm}\addtocounter{footnote}{1}\addtocounter{footnote}{-1} \\
  \hline
  \begin{minipage}{1.1cm}
  Dispersive \\ Delay \\ Structure \\ (DDS)
  \end{minipage}
  & \hspace{-0.4cm}\begin{minipage}{4cm}
  \begin{itemize}
    \item emphasizes that $\tau$ is \\ frequency-dependent
    \item sometimes used in \\ photonics, surface \\ acoustic waves and \\ magnetostatic waves
  \end{itemize}
  \end{minipage}
    & \hspace{-0.4cm}\begin{minipage}{3.3cm}
  \begin{itemize}\vspace{1mm}
    \item long $\Rightarrow$ acronym DDS
    \item DDS also stands for Direct Digital \\ Synthesizer
    \item not general \\ (only structure)
    \end{itemize}\vspace{1mm}
  \end{minipage} \\
  \hline
  \begin{minipage}{1.1cm}
  All-pass \\ Network
  \end{minipage}
  & \hspace{-0.4cm}\begin{minipage}{4cm}
  \begin{itemize}
    \item used for a long \\ time in microwaves
  \end{itemize}
  \end{minipage}
  & \hspace{-0.4cm}\begin{minipage}{3.3cm}
  \begin{itemize}\vspace{1mm}
    \item not all phasers are \\ all-pass (e.g.~\cite{Zhang_TMTT_03_2013})
    \item confusion with \\ dispersion-less (linear \\ phase) all-pass
    \item not general \\ (only network)
  \end{itemize}\vspace{1mm}
  \end{minipage} \\
  \hline
  \begin{minipage}{1.1cm}
  Arbitrary Phase \\ Network
  \end{minipage}
  & \hspace{-0.4cm}\begin{minipage}{4cm}
  \begin{itemize}
    \item emphasis on phase \\ vs magnitude
  \end{itemize}
  \end{minipage}
    & \hspace{-0.4cm}\begin{minipage}{3.3cm}
  \begin{itemize}\vspace{1mm}
    \item phase is designable \\ but not arbitrary
    \item not general \\ (only network)
    \item confusion with \\ phase shifters \\ (single frequency)
  \end{itemize}
  \end{minipage} \\
\end{tabular}
\label{tab:phaser_term}
\end{table}

\begin{figure}[h!]
\centering
\includegraphics[width=0.35\textwidth,page=2]{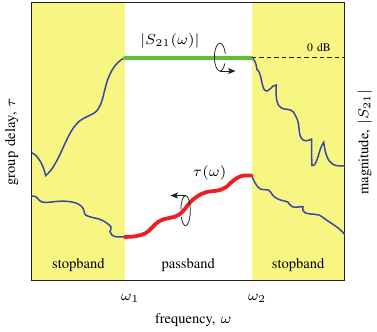}
\caption{Ideal phaser design: arbitrary group delay with flat and lossless magnitude response over a given frequency band.}
\label{fig:phaser_eng}
\end{figure}

An \emph{ideal} phaser is a phaser that would exhibit an arbitrary group delay with flat and lossless magnitude over a given frequency band, as illustrated in Fig.~\ref{fig:phaser_eng}. Obviously, such an ideal response is practically unrealizable. Realizing a response as close as possible to this ideal one for specific applications is the objective of phaser synthesis, which will be discussed in Sec.~\ref{sec:phaser_synthesis}.

The phase versus frequency function of a phaser, \emph{within a certain specified bandwidth} may be expanded in a Taylor series around a frequency (often the center frequency) within the phaser's bandwidth, $\omega_0$:
\begin{subequations}\label{eq:phase_Taylor}
\begin{equation}\label{eq:phase_Taylor_ser}
\phi(\omega)
= \phi_0
+\phi_1(\omega-\omega_0)
+\dfrac{\phi_2}{2}(\omega-\omega_0)^2
+\dfrac{\phi_3}{6}(\omega-\omega_0)^3+
\ldots
\end{equation}
\begin{equation}
\text{with}\quad\phi_k=\left.\dfrac{\partial^k\phi(\omega)}{\partial\omega^k}\right|_{\omega=\omega_0}.
\end{equation}
\end{subequations}
\noindent The first three coefficients of the series explicitly read
\begin{subequations}
\begin{equation}
\phi_0=\phi(\omega_0)\quad\text{(rad)},
\end{equation}
\begin{equation}\label{eq:phi1}
\phi_1=\left.\dfrac{\partial\phi}{\partial\omega}\right|_{\omega_0}
=-\tau(\omega_0)\quad\text{(s)},
\end{equation}
\begin{equation}\label{eq:phi2}
\phi_2=\left.\dfrac{\partial^2\phi}{\partial\omega^2}\right|_{\omega_0}
=\left.\dfrac{\partial\tau}{\partial\omega}\right|_{\omega_0}
=s_\tau(\omega_0)\quad\text{(s$^2$/rad)},
\end{equation}
\end{subequations}
\noindent and represent the \emph{phase, group delay and  group delay dispersion parameters} of the phaser at $\omega_0$, respectively. The group delay function of the phaser is then obtained from~\eqref{eq:phase_Taylor_ser} as
\begin{equation}\label{eq:delay_Taylor}
\tau(\omega)
=-\dfrac{\partial\phi(\omega)}{\partial\omega}
=-\phi_1
-\phi_2(\omega-\omega_0)
-\dfrac{\phi_3}{2}(\omega-\omega_0)^2-
\ldots,
\end{equation}
where one notes, consistently with \eqref{eq:phi2}, that $\phi_2$ is the \emph{slope of the group delay function at $\omega_0$}.

\indent The most common phaser group delay response is the \emph{linear group delay} one. This response corresponds to a second-order nonlinear phase function, where $\phi_2\neq 0$ \mbox{($\phi_2\gtrless 0$)}, $\phi_0$ being an arbitrary phase reference and $\phi_1>0$ depending on the electrical length of the phaser, and $\phi_k=0$ for $k>2$ in~\eqref{eq:phase_Taylor} and \eqref{eq:delay_Taylor}. It is the response used in Fig.~\ref{fig:ASP_basic_effects}(a) with $\phi_2>0$, the response required for real-time Fourier transforming, to be presented in Sec.~\ref{sec:RTFT}, and this is the function that will be considered in most of the examples of this paper. However, other group delay functions may be required, such as for instance the stepped group delay function, that is shown in Fig.~\ref{fig:ASP_basic_effects}(b) for the case of two steps and that will be used in the spectrum sniffer to be presented in Sec.~\ref{sec:ASP_apps}.

The realization of phasers requires controllable dispersion for efficient group delay engineering. It may follow two distinct but possibly combined~\cite{Gupta_MOTL_03_2012} approaches: the \emph{medium approach} and the \emph{network approach}. The medium approach consists in exploiting the natural dispersion of highly dispersive materials~\cite{Saleh_Teich_FP,Ashcroft_book,Jackson_book_CED} or the artificial dispersion of metamaterial structures~\cite{Caloz_Wiley_2006,Carignan_TMTT_10_2011,Caloz_PIEEE_10_2011} for phasing. It essentially synthesizes constitutive parameters, which are most generally bi-anisotropic~\cite{Kong_book} and which correspond to the fundamental space harmonic in the case of periodic metamaterials~\cite{Caloz_MT_2009}. This approach is particularly indicated when spatial dispersion or/and radiation is involved in addition to temporal dispersion, as will be shown in the case of the real-time Fourier transformer to be presented in Sec.~\ref{sec:ASP_apps}. The network approach consists in using microwave-type coupler and filter structures and techniques~\cite{Guillemin_SOB_1957},\cite{MYJ_AHP_1980,Cameron_Kudsia_Mansour_MFCS} for phasing. These structures are most often aperiodic. They offer a higher degree of temporal dispersion control compared to material or metamaterial phasers, as will be shown in Sec.~\ref{sec:phaser_synthesis}, and they are therefore preferred for purely wave-guiding phasers.

\begin{figure}[h!]
\centering
\includegraphics[width=0.5\textwidth,,page=3]{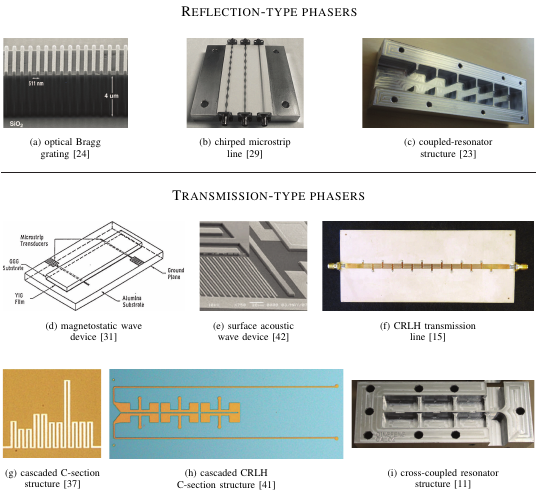}
\caption{Some of the most common microwave phasers.}
\label{fig:phasers}
\end{figure}

Figure~\ref{fig:phasers} shows some of the most common microwave phasers reported to date. These phasers can be broadly classified as \emph{reflection-type phasers and transmission-type phasers}, some of the formers being shown in Figs.~\ref{fig:phasers}(a)-(c) and some of the latter being shown in Figs.~\ref{fig:phasers}(d)-(i). Reflection-type phasers require a circulator or a directional coupler at their input to provide the required functional two-port network response suggested in Fig.~\ref{fig:ASP_basic_effects}~\cite{Zhang_TMTT_08_2012}, while transmission-type phasers are inherently two-port network components and may be therefore straightforwardly integrated into ASP system.

\emph{Bragg grating} phasers are realized by cascading Bragg grating sections~[Fig.~\ref{fig:phasers}(a)] of progressively varying periods, so as to reflect different frequencies at different locations and hence with different delays~\cite{Saleh_Teich_FP,nanoweb_grating,Kashyap_FBG}. Bragg grating phasers are excessively bulky at low microwave frequencies, but might be sometimes replaced by non-uniform artificial substrate planar structures based on the same principle~\cite{Coulombe_TMTT_08_2009}.
\emph{Chirped microstrip line} phasers [Fig.~\ref{fig:phasers}(b)] represent another planar technology with position-frequency dependent sections, with the benefit of some extra design flexibility provided by smooth discontinuities but the drawback of relatively large size, and hence, high loss~\cite{Laso_MWCL_12_2001,Laso_TMTT_03_2003,Schwartz_MWCL_04_2006}.
\emph{Coupled-resonator structure} phasers~[Fig.~\ref{fig:phasers}(c)], to be discussed next, are more compact (in a given technology, e.g. planar or waveguide) and can exactly follow specified group delay responses~\cite{Zhang_TMTT_08_2012}, as will be shown in Sec.~\ref{sec:phaser_synthesis}.
\emph{Magnetostatic wave (MSW)} phasers~[Fig.~\ref{fig:phasers}(d)] utilize the inherent dispersion of quasi-static modes in ferrimagnetic films, which may be controlled via layer, strip and boundary structural parameters~\cite{Ishak_PIEE_02_1988,Adam_MW_1991,Stancil_TMW_1993}. Due to the quasi-static nature of magnetostatic modes, the MSW wavelength is about four orders of magnitude smaller than that of electromagnetic modes, which leads to very compact devices. MSW technology was very popular in the 1980ies but it is nowadays rarely used, mostly due to the requirement of a biasing magnet.
\emph{Surface acoustic wave (SAW)} phasers~[Fig.~\ref{fig:phasers}(e)] owe their dispersive properties to the distribution of electrodes on piezoelectric substrates~\cite{Campbell_SAW_1989,Lewis_SAW_OSP_2005}. They are even more compact than MSW devices, the relevant acoustic wavelength being over six orders of magnitude smaller than that of electromagnetic waves. SAW devices are abundantly used in the microwave industry but they are mostly restricted to frequencies below the X~band, due to material limitations, and they are inapplicable to millimeter-wave and terahertz frequencies.
Composite right/left-handed (CRLH) transmission lines~[Fig.~\ref{fig:phasers}(f)] and other metamaterial-type (sub-wavelength unit-cell) artificial transmission lines may have their dispersion controlled in terms of the Taylor coefficients of their wavenumber, under the constraint of fixed Bloch impedance for broad-band matching~\cite{Caloz_Wiley_2006,Caloz_PIEEE_10_2011,Gupta_TMTT_04_2009,Abielmona_TMTT_11_2009}. They are less compact than MSW-based and SAW-based phasers, but they may be scaled to virtually any frequency, feature size-bandwidth independent characteristic, due to their lumped unit cell, and are particularly suited to radiative and spatially dispersive applications~\cite{Caloz_PIEEE_10_2011}.
\emph{Cascaded C-section structure} phasers~[Fig.~\ref{fig:phasers}(g)], to be discussed next, offer the benefit of higher compactness and, as coupled-resonator phasers, can exactly follow specified group delay responses~\cite{Gupta_TMTT_09_2010,Gupta_IJCTA_0000,Zhang_IJRMCAE_TBP}, as will be shown in Sec.~\ref{sec:phaser_synthesis}.
\emph{Cascaded CRLH C-section structure} phasers incorporate CRLH C-sections~[Fig.~\ref{fig:phasers}(h)] so as to benefit from the enhanced coupling characteristics of CRLH coupled-line couplers~\cite{Nguyen_TMTT_5_2007} for higher dispersion and broader bandwidth~\cite{Gupta_TMTT_12_2012}.
Finally, \emph{cross-coupled resonator structure} phasers~[Fig.~\ref{fig:phasers}(i)] exhibit the same configuration types as cross-coupled filters~\cite{Cameron_Kudsia_Mansour_MFCS}. As in filters, the addition of cross-coupling to sequential coupling provides extra degrees of freedom in phasers, and the corresponding phaser responses can follow exact specifications~\cite{Zhang_TMTT_03_2013}, as will be shown in~Sec.~\ref{sec:phaser_synthesis}.

\begin{figure}[h!]
\centering
\includegraphics[width=0.5\textwidth,page=4]{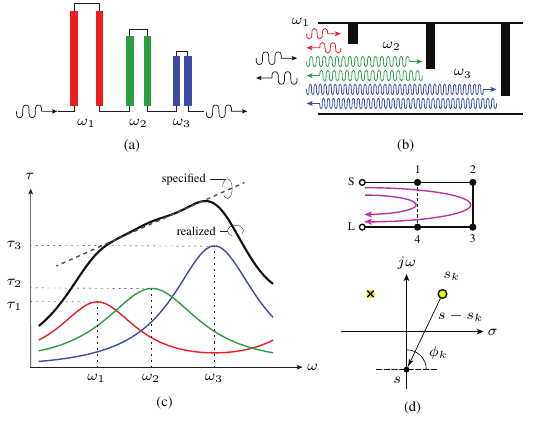}
\caption{Principles of the selected phasers. (a)~Cascaded C-section phaser [Fig.~\ref{fig:phasers}(g)]. (b)~Coupled-resonator phaser [Fig.~\ref{fig:phasers}(c)]. (c)~Group delay response formation for (a) and (b). (d)~Cross-coupled phaser [Fig.~\ref{fig:phasers}(i)].}
\label{fig:network_phasers}
\end{figure}

Based on the above comparisons, coupled-resonator phasers [Fig.~\ref{fig:phasers}(c)], cascaded C-section phasers [Fig.~\ref{fig:phasers}(g)-(h)] and cross-coupled resonator phasers [Fig.~\ref{fig:phasers}(i)], which are all network-type phasers, seem the most promising technologies for ASP systems restricted to guided-wave requirements, while artificial (e.g. CRLH) transmission line phasers seem the best option for radiated and spatially dispersive phasers. The latter have been largely covered in recent literature. We shall therefore focus here on the former. Let us first explain their basic principle with the help of Fig.~\ref{fig:network_phasers}.

Figure~\ref{fig:network_phasers}(a) represents a three-section cascaded C-section phaser as a simplified version of the phaser shown in Fig.~\ref{fig:phasers}(g). A C-section is a coupled-line coupler with two end ports interconnected so as to form a two-port network. It is an \emph{all-pass} network \mbox{($|S_{21}(\omega)|=1,\forall\omega$)}, whose transfer function reads \mbox{$S_{21}(\theta)=(a-j\tan\theta)/(a+j\tan\theta)=e^{j\phi}$} with \mbox{$a=\sqrt{(1+k)/(1-k)}$}, where $\theta=\beta\ell$ is the electrical length of the structure, $\beta$ and $\ell$ being its wavenumber and physical length, respectively, and $k$ is the coupler's coupling coefficient~\cite{Steenart_TMTT_01_1963}. The corresponding group delay is then \mbox{$\tau(\omega)=-d\phi/d\omega=\left\{2a/[1+(a^2-1)\cos^2\theta]\right\}d\theta/d\omega$}, and is seen to reach maxima at $\theta=(2m+1)\pi$ ($m$ integer) or $\ell=(2m+1)\lambda/4$, the first maximum being at $\ell=\lambda/4$, as represented in the color curves of Fig.~\ref{fig:network_phasers}(c) for the different C-section lengths. A detailed wave-interference explanation for the C-section's response is provided in~\cite{Gupta_TMTT_12_2012}. Cascading C-sections of different lengths forms then a \emph{transmission-type} network, whose total group delay at each frequency is the sum of the delays incurred by each C-section, as represented by the black curve in~Fig.~\ref{fig:network_phasers}(c). A phaser following a prescribed group delay function can then be realized by controlling the sizes of the different C-sections, their coupling coefficient and their number, as will be shown in Sec.~\ref{sec:phaser_synthesis}.

The coupled-resonator phaser represented in Fig.~\ref{fig:network_phasers}(b), corresponding to the component shown in Fig.~\ref{fig:phasers}(c), is an \emph{all-pass} network, as the previous phaser, but it is of \emph{reflection-type}. Its operation is essentially similar to that of a Bragg grating~[Fig.~\ref{fig:phasers}(a)] phaser, except that its response is achieved by aperture-resonator pairs rather than by a modulated periodic lattice, which provides higher design flexibility, allows perfect synthesis accuracy and leads to much more compact devices. Intuitively, the lowest frequency components of the input signal are reflected at the largest apertures (diaphragms or irises in a waveguide) whereas the highest frequencies are reflected at the smallest apertures, as illustrated in Fig.~\ref{fig:network_phasers}(b), which leads to the desired total delay versus frequency curve in black in~Fig.~\ref{fig:network_phasers}(c) with proper synthesis. Compared to the cascaded C-section phaser, this phaser has the drawback of requiring a circulator or a coupler, since it is a reflection-type as opposed to a transmission-type component. However, it provides higher dispersion (i.e. higher~$|\phi_2|$) for narrow-band applications, as will be illustrated in the results of Sec.~\ref{sec:phaser_synthesis} and shown from first principles elsewhere. It may be realized in planar technology as well as in waveguide technology.

Finally, the cross-resonator phaser represented in Fig.~\ref{fig:network_phasers}(d), corresponding to the component shown in Fig.~\ref{fig:phasers}(i), is the latest one investigated by the authors~\cite{Zhang_TMTT_03_2013}. Being of transmission type, it does not require a circulator or coupler, as the cascaded C-section phaser, while, exploiting cross-coupling, it is likely to provide maximal resolution after optimization. Its operation principle is less intuitive than that of the other two phasers, but may be understood with the help of Fig.~\ref{fig:network_phasers}(d). In a cross-coupled \emph{filter}, the different wave paths are typically designed so as to produce destructive interference at the frequencies of desired attenuation. The transmission zeros, $s_k$, of the transfer function, which may be generally written \mbox{$S_{21}=\Pi_k(s-s_k)/H(s)$} where $H(s)$ is a Hurwitz polynomial~\cite{Cameron_Kudsia_Mansour_MFCS}, are then placed on the imaginary axis of the complex~$s$ plane. As $s$ spans the imaginary axis, corresponding to real frequencies ($s=j\omega$), it crosses the transmission zeros, which builds the \emph{magnitude} response at the cutoff and in the stopband. In contrast, in a cross-coupled \emph{phaser}, the transmission zeros are not intended to generate attenuation or stop-bands but to shape the \emph{phase} response. Therefore, they are placed \emph{off the imaginary axis} and they are paired with transmission poles, as illustrated at the bottom of Fig.~\ref{fig:network_phasers}(d), which cancels the attenuation effect while ensuring a phase effect. Note that this phaser is not an all-pass but a \emph{band-pass} phaser, where the presence of the stopbands will relax the passband constraints for more flexible phaser design, as will be shown in~Sec.~\ref{sec:phaser_synthesis}.

The cancelation by transmission poles of transmission zeros of the transfer function placed off the imaginary axis of the complex frequency plane is characteristic of all-pass functions. This effect may be verified as follows in the case of cascaded C-section phasers. Substituting the frequency mapping function
\begin{equation*}
s=j\tan\theta
\end{equation*}
into the C-section transfer function~\cite{Gupta_IJCTA_0000,Gupta_TMTT_12_2012,Steenart_TMTT_01_1963}
\begin{equation*}
S_{21}(\theta)=\frac{a-j\tan\theta}{a+j\tan\theta}
\end{equation*}
yields
\begin{equation*}
S_{21}(s)=-\frac{s-a}{s+a}.
\end{equation*}
\indent In the last relation, the transmission zero, \mbox{$s=a$}, and the transmission pole, $s=-a$, are here both real and are placed anti-symmetrically with respect to the imaginary axis, which leads, upon the substitution $s=j\omega$, to the all-pass response, $|S_{21}|=1,\forall\omega$.\\
\indent In a \emph{D-section}, which represents the next higher-order all-pass function above the C-section, the two transmission zeros and poles are both complex and are placed symmetrically with respect to the imaginary axis~\cite{Zhang_IJRMCAE_TBP,Steenart_TMTT_01_1963}, as represented in Fig.~\textcolor{red}{\ref{fig:network_phasers}}(d), which also ensures an all-pass response.

\section{Real-Time Fourier Transforming}\label{sec:RTFT}

The Fourier transform~\cite{Papoulis_FIA} is one of the most fundamental operations in science and technology. In today's microwave systems, it is most often performed digitally using the fast-Fourier transform~(FFT) algorithm~\cite{Oppenheim_Schafer_DTSP}, and it represents for instance the basis of orthogonal frequency division multiplexing~(OFDM) in wireless communications~\cite{Tse_Viswanath_FWC_2005}. In the realm of ASP, Fourier transformation may be performed in real time and is then called \emph{real-time Fourier transformation}~(RTFT)~\cite{Jannson_OL_04_1983}. We shall describe here the principle of RTFT, although it is not directly used in many ASP applications, because it highlights some fundamental characteristics of ASP systems, that will be further detailed in terms of phasers in Sec.~\ref{sec:char_enhanc} and that will subsequently guide synthesis specifications in Sec.~\ref{sec:phaser_synthesis}.

\begin{figure}[h!]
\centering
\includegraphics[width=0.45\textwidth,page=5]{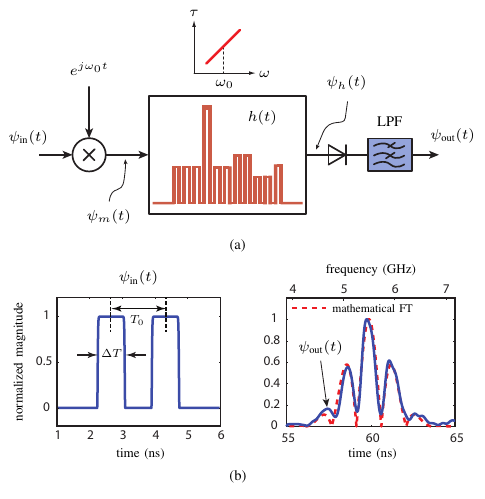}
\caption{Real-time Fourier transformer~(RTFT). (a)~Microwave implementation. (b)~Result example using a measured cascaded C-section phaser of the type of Fig.~\ref{fig:phasers}(g).}
\label{fig:RTFT}
\end{figure}

Figure~\ref{fig:RTFT}(a) shows the block diagram of a microwave RTFT system. RTFT requires a linear group delay (or quadratic phase) phaser response, as represented in the figure. The impulse response of such a phaser, $h(t)$, is derived in the Appendix, where it is given by Eq.~\eqref{eq:impulse_resp}. As shown in Fig.~\ref{fig:RTFT}(a), the base-band input signal, $\psi_\text{in}(t)$, to be Fourier-analyzed in the time domain, is first up-converted to the microwave frequency band of the phaser, which is centered at $\omega_0$, to yield \mbox{$\psi_m(t)=\psi_\text{in}(t)e^{j\omega_0t}$}. This modulated signal is then passed through the phaser, which is considered here linear and whose output response, $\psi_h(t)$, is hence the convolution of $\psi_m(t)$ with $h(t)$. On has thus, using~\eqref{eq:impulse_resp}:
\begin{equation}\label{eq:RTFT_deriv}
\begin{split}
\psi_h(t) &= [\psi_\text{in}(t)e^{j\omega_0t}]\ast h(t)\\
&=\int_{-\infty}^{+\infty}\left[\psi_\text{in}(\tau) e^{j\omega_0\tau}\right]\left[\gamma e^{-j\frac{\beta (t-\tau)}{\phi_2}}e^{-j\frac{(t-\tau)^2}{2\phi_2}}\right]
d\tau\\
&=\gamma e^{-j\frac{\beta t}{\phi_2}}e^{-j\frac{t^2}{2\phi_2}}\int_{-\infty}^{+\infty}\psi_\text{in}(\tau) e^{j\omega_0\tau}e^{j\frac{\beta\tau}{\phi_2}}e^{j\frac{t\tau}{\phi_2}}e^{-j\frac{\tau^2}{2\phi_2}}d\tau\\
&=\gamma e^{-j\frac{\beta t}{\phi_2}}e^{-j\frac{t^2}{2\phi_2}}\int_{-\infty}^{+\infty}\psi_\text{in}(\tau) e^{j\frac{(\phi_1+t)\tau}{\phi_2}}e^{-j\frac{\tau^2}{2\phi_2}}d\tau.
\end{split}
\end{equation}

\indent In the last expression of~\eqref{eq:RTFT_deriv}, the second exponential of the integrand reduces to unity if $\tau_\text{max}^2/(2|\phi_2|)\ll\pi$, or~\cite{Muriel_OL_01_1999}
\begin{equation}\label{eq:RTFT_cond}
\dfrac{T_{\psi_\text{in}}^2}{2\pi|\phi_2|}\ll 1,
\end{equation}
where $T_{\psi_\text{in}}$ represents the duration of the input signal~(assumed to extend from $t=0$ to $t=T_{\psi_\text{in}}$), which limits the span of the integral~\eqref{eq:RTFT_deriv}. Note that condition~\eqref{eq:RTFT_cond} does \emph{not} eliminate the first exponential of the integrand in~\eqref{eq:RTFT_deriv}, despite the presence of the~$\phi_2$ factor in its argument, because the numerator of its argument is enhanced by the group delay term $\phi_1$ and is multiplied by~$t$, which is much greater than $\tau$ over most of the duration of $\psi_h(t)$, due to the assumed large dispersion~($\phi_2$) in~\eqref{eq:RTFT_cond}. In the interval of $\psi_h(t)$'s duration, we have $t\sim\tau$, but $\phi_1$ may still ensure the dominance of the first exponential of the integrand thanks to the term $\phi_1$.

Under condition~\eqref{eq:RTFT_cond}, the absolute value of~\eqref{eq:RTFT_deriv} reduces then to the output signal of Fig.~\ref{fig:RTFT}(a), after envelope detection,
\begin{subequations}\label{eq:final_RTFT_expr}
\begin{equation}\label{eq:FT_RTFT}
\psi_\text{out}[\omega(t)]
=|\psi_h(t)|
=\sqrt{\dfrac{2\pi}{\phi_2}}
\int_{-\infty}^{+\infty}\psi_\text{in}(\tau) e^{j\omega(t)\tau}d\tau
\end{equation}
with
\begin{equation}\label{eq:map_RTFT}
\omega(t)=\dfrac{\phi_1+t}{\phi_2}.
\end{equation}
\end{subequations}
The signal~$\psi_\text{out}[\omega(t)]$ in~\eqref{eq:FT_RTFT} is essentially the Fourier transform of $\psi_\text{in}(t)$ \emph{in the time domain}: its shape versus time is identical to the Fourier transform of $\psi_\text{in}(t)$, whose actual frequencies are obtained by the mapping function~\eqref{eq:map_RTFT}. Figure~\ref{fig:RTFT}(b) shows an example of an RTFT result involving a measured cascaded C-section phaser.

\section{Phaser Characteristics and Enhancement}\label{sec:char_enhanc}

The performance of an ASP system directly depends on the characteristics of the phaser that it utilizes. The three most important phaser characteristics in this regard are the \emph{resolution}, the \emph{absolute bandwidth} and the \emph{magnitude balance}.

The resolution characteristic can be understood intuitively with the help of Fig.~\ref{fig:ASP_basic_effects}(b), which represents in essence a \emph{frequency discriminator}~\cite{Stewart_IRE_01_1952}, one of the most basic ASP systems~\cite{Nikfal_TMTT_06_2011}. If the time difference, $\Delta\tau$, between the group delays of the two input modulation frequencies, $\omega_{01}$ and $\omega_{02}$, is less than the duration $T_0$ of the input pulse, the two pulses at the output are not fully separated in time, i.e. \emph{resolved}, and can therefore not be unambiguously discriminated. Then the frequencies $\omega_{01}$ and $\omega_{02}$ cannot be clearly detected in time: the resolution of the ASP system is insufficient. This indicates that the \emph{resolution}, or capability of a phaser to resolve the frequency components of a signal, is proportional to the group delay difference (or swing), $\Delta\tau$. At the same time, if $T_0$ is reduced, then $\Delta\tau$ can also be reduced without loosing discrimination; the shorter $T_0$ is, the smaller $\Delta\tau$ can be. So, the resolution of an ASP \emph{system} is also inversely proportional to duration of its input pulse, $T_0$. So, the resolution, that we shall denote $\varrho$, should follow $\varrho\propto\Delta\tau/T_0$. But $T_0$ is inversely proportional to the pulse bandwidth, $B_0$, so that $\varrho\propto\Delta\tau B_0$. Now, the spectrum of the pulse must clearly be fully contained in the bandwidth of the phaser, $|\Delta\omega|$, or $B_0\leq|\Delta\omega|$, so that we also have $B_0\propto|\Delta\omega|$. Finally, the ASP resolution of a \emph{phaser} may thus be defined as the unit-less quantity
\begin{equation}\label{eq:phaser_resol}
\varrho=|\Delta\tau|\cdot|\Delta\omega|,
\end{equation}
\noindent where $\Delta\tau$ represents the group delay swing provided by the phaser over its frequency bandwidth $\Delta\omega$.

It is instructive to consider the particular case of a linear group delay phaser at this point. This is case considered in the system of Fig.~\ref{fig:ASP_basic_effects}(a), which may operate as a \emph{frequency meter}~\cite{Steen_TBE_03_1979} where the unknown frequency to be measured, $\omega_0$, is found from the time position of the maximum of the output chirped pulse using the phaser law, $\tau(\omega)$. In this case, from \eqref{eq:phi2}, we have \mbox{$\phi_2=\partial\tau/\partial\omega|_{\omega_0}=\Delta\tau/\Delta\omega$}, and using this relation to eliminate $\Delta\tau$ in~\eqref{eq:phaser_resol}, yields
\begin{equation}\label{eq:phaser_resol_lin}
\varrho_\text{lin}=|\phi_2|\cdot|\Delta\omega|^2.
\end{equation}
This equation reveals that the resolution is proportional to the group delay dispersion parameter, $|\phi_2|$ ($\phi_2>0$ (resp. \mbox{$\phi_2<0$}) for a positive (resp. negative) group delay slope) and to the square of the phaser bandwidth. It is fully consistent with the RTFT condition~\eqref{fig:RTFT}, both directly in terms of $|\phi_2|$ and indirectly in terms of $\Delta\omega\propto 1/T_0=1/T_{\psi_\text{in}}$, leading to $\varrho_\text{lin}=|\phi_2|/|T_{\psi_\text{in}}|^2$: the accuracy of RTFT is proportional to $|\phi_2|$, or equivalently, for a given accuracy, increasing $|\phi_2|$ allows one to analyze signals with larger duration, $T_{\psi_\text{in}}$.

\begin{figure}[h!]
\centering
\includegraphics[width=0.45\textwidth,page=6]{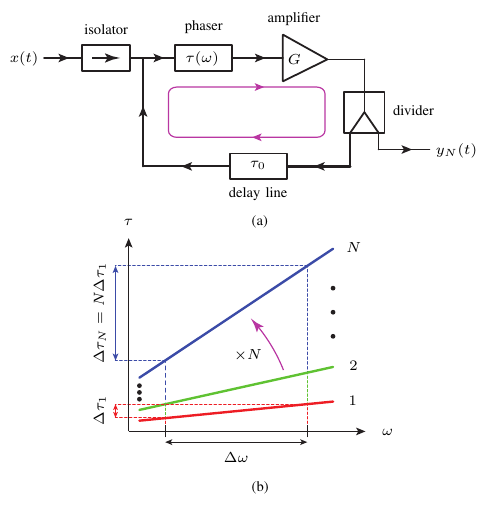}
\caption{Feedback loop resolution enhancement principle~\cite{Nikfal_TMTT_06_2011}. (a)~Circuit block diagram. (b)~Group delay slope~($|\phi_2|$) multiplication.}
\label{fig:loop}
\end{figure}

If the ASP resolution of a given phaser (e.g. any phaser in Fig.~\ref{fig:phasers}) is insufficient, it may be enhanced by using the feedback loop circuit shown in Fig.~\ref{fig:loop}(a)~\cite{Nikfal_TMTT_06_2011}. In this system, the signal to be processed, $x(t)$, is passed $N$ times through a feedback loop including the phaser plus and amplifier, regenerating the signal level, and a constant delay line, avoiding signal self-overlapping. The group delay slope is progressively increased as the signal loops in the system, as shown in Fig.~\ref{fig:loop}(b). After the second pass, all the delays experienced by the different frequency components of $x(t)$ have been doubled, and hence the slope has been doubled. After the $N^\text{th}$ pass, the slope has been multiplied by $N$, and thus, the bandwidth $\Delta\omega$ being unchanged, the resolution as defined by~\eqref{eq:phaser_resol} or~\eqref{eq:phaser_resol_lin}  has been multiplied by the same factor:~$\varrho_N=N\varrho_1=N|\phi_2|$. The processed signal, $y_N(t)$, is then extracted from the loop using an appropriate counting-switching scheme. A brute-force resolution enhancement approach would consist in cascading $N$ times the same phaser, with the amplifier at the output of the chain. However, the resulting system would have prohibitive drawbacks: i)~it would be excessively large; ii)~it would suffer from high insertion loss due to multiple-reflection mismatch between the different phasers; iii)~it would have a poor signal-to-noise due to the low level of the signal reaching the amplifier.

In applications where the relevant frequency bandwidth is very small, such as for instance in Doppler radar~\cite{Skolnik_book_1990} or in narrow-band signal RTFT, the \emph{absolute frequency bandwidth} is another essential figure of merit. Several phasers, such as the transmission-type C-section phaser to be quantitatively characterized in Sec.~\ref{sec:phaser_synthesis}, provide a group delay swing in the order of~1~ns over a bandwidth in the order of~1~ns, corresponding to the relatively high resolution of $\varrho=1$, according to~\eqref{eq:phaser_resol}. However, if the relevant frequency band is much smaller~(e.g.~10~MHz), the corresponding \emph{relevant} resolution, corresponding to this frequency band, is much smaller~\mbox{(e.g.~1~ns~$\cdot$~1~MHz$~=0.01$)}, and may be largely insufficient. In this case, another phaser technology must be used, such as for instance that of reflection-type coupled-resonator phasers, which offer an even higher group delay swing over a much smaller bandwidth, as will be quantitatively exemplified in~Sec.~\ref{sec:phaser_synthesis}.

Let us now consider the third and last phaser characteristic, magnitude balance. Magnitude balance is defined as a flat transmission magnitude, $|S_{21}|$, over the frequency band of the phaser, as represented in Fig.~\ref{fig:phaser_eng}. In contrast to what is shown for the ideal case represented in Fig.~\ref{fig:phaser_eng}, the magnitude cannot be of 0~dB, due to material and structural losses, but it may be amplified, as in any processing system, if required. However, magnitude balance must be achieved as well as possible to ensure high ASP performance. We shall next see how magnitude imbalance comes about, how it affects ASP, and how to remedy it for higher ASP performance.

\begin{figure}[h!]
\centering
\includegraphics[width=0.5\textwidth,page=7]{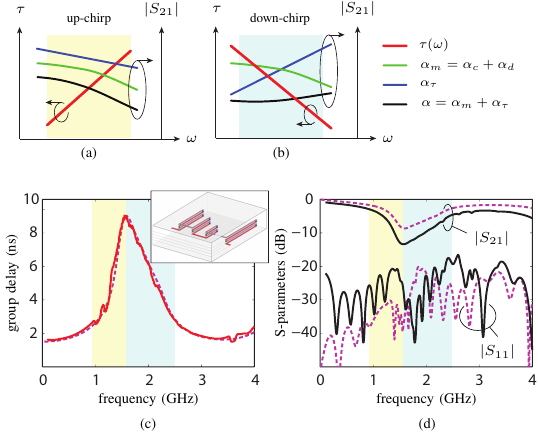}
\caption{Magnitude imbalance due to loss in an unequalized phaser. (a)~Trends of material and dispersion losses for an up-chirp phaser. (b)~Same for a down-chirp phaser. (c)~Group delay of the multilayer C-section phaser shown in the inset (solid: measurement, dashed: full-wave)~\cite{Horii_MWCL_01_2012}. (d)~Corresponding S-parameter magnitudes.}
\label{fig:mag_imbalance}
\end{figure}

The problematic of magnitude imbalance is described in Fig.~\ref{fig:mag_imbalance}. Magnitude imbalance, i.e. a frequency varying transmission magnitude, $|S_{21}|$, in a phaser is caused by loss. The main loss contributors in a network phaser of effective length $\ell$ are material loss, $e^{-\alpha_m(\omega)\ell}$, including conductor loss, $e^{-\alpha_c(\omega)\ell}$, and dielectric loss, $e^{-\alpha_d(\omega)\ell}$, although radiation loss may also play a significant role in open structures at high frequencies. It is well-known that these losses increase with increasing frequencies in any microwave structure~\cite{Pozar_ME_2011}, as schematically represented by the green curves in Figs.~\ref{fig:mag_imbalance}(a) and (b), which represent the responses of a positive linear chirp (up-chirp) phaser and a negative linear chirp (down-chirp) phaser, respectively. However, an effect that is specific to phasers must also be considered, the \emph{phasing loss effect}. The amount of dissipation produced by any type of loss mechanism is necessarily proportional to the amount of time that the signal spends in the system. In a phaser, this time greatly varies across the phaser bandwidth, and this variation is even desired to be relatively large for high resolution, as seen in \eqref{eq:phaser_resol}. Therefore, phasing loss, which may be modeled by the attenuation factor $e^{-\alpha_\tau(\omega)\ell}$, strongly depends on frequency, and increases (resp. decreases) with frequency in an up-chirp (resp. down-chirp) phaser, as represented by the blue curves in Fig.~\ref{fig:mag_imbalance}(a) [resp.~Figs~\ref{fig:mag_imbalance}(b)]. As a result, an unequalized up-chirp phaser will feature a negative-slope magnitude imbalance, as represented by the black curve in Fig.~\ref{fig:mag_imbalance}(a), while an unequalized down-chirp phaser may exhibit a negative-slope or a positive-slop magnitude depending which is the dominant effect between $e^{-\alpha_\tau(\omega)\ell}$ and $e^{-\alpha_m(\omega)\ell}$. Figures~\ref{fig:mag_imbalance}(c) and (d) show the group delay and magnitude responses, respectively, of a practical phaser including a both a positive slope (lower frequency range) and a negative slope (upper frequency range) of the group delay, each of which are independently utilizable, to illustrate this effect.

\begin{figure}[h!]
\centering
\includegraphics[width=0.35\textwidth,page=8]{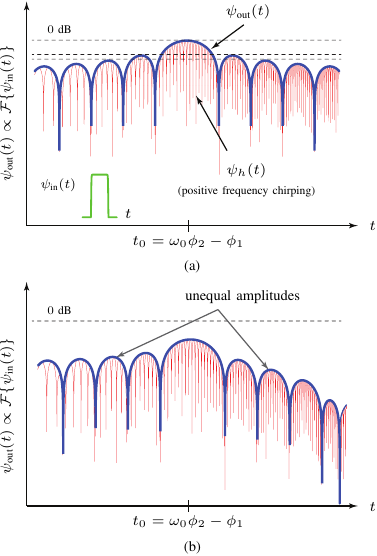}
\caption{Effect of magnitude imbalance in the case of real-time Fourier transformation~[Fig.~\ref{fig:RTFT} and Eq.~\eqref{eq:final_RTFT_expr}]. (a)~Lossless Fourier transform (log scale) of the rect signal shown in the inset. (b)~Corresponding lossy transform with negative slope magnitude imbalance and exaggerated loss.}
\label{fig:FT_mag_unb_effect}
\end{figure}

Phaser magnitude imbalance unbalances ASP systems in various fashions. For instance, in the frequency discriminator of Fig.~\ref{fig:ASP_basic_effects}(b), the pulse associated with the lossier frequency may fall below the noise floor. As an other example, Fig.~\ref{fig:mag_imbalance} illustrates the effect of magnitude imbalance in an RTFT system~(Sec.~\ref{sec:RTFT}) for a rectangular input base-band pulse. In an ideal system, the output signal would be a perfect sinc function of time, properly representing the Fourier transform of the input rectangular signal according to~\eqref{eq:final_RTFT_expr}, with the center frequency, $\omega_0$, appearing at \mbox{$t_0=\omega_0\phi_2-\phi_1$} according to~\eqref{eq:map_RTFT}. This is the situation is represented in Fig.~\ref{fig:FT_mag_unb_effect}(a). In the case of the lossy transform, represented in Fig.~\ref{fig:FT_mag_unb_effect}(b), magnitude imbalance has distorted the Fourier transform result.

Magnitude imbalance can be remedied using various techniques. A possible technique would consist in using a resistive magnitude equalization network, having negligible effect on the phase response and sacrificing on the transmission level of the less lossy parts of the spectrum~\cite{Fejzuli_HFE_06_2006}. A more subtle technique is to account for the frequency-dependent losses in the phaser synthesis, which is possible in cross-coupled resonator phasers, to be presented in Sec.~\ref{sec:phaser_synthesis}.

\section{Phaser Synthesis}\label{sec:phaser_synthesis}

As mentioned in Sec.~\ref{sec:core_ASP_phaser}, the objective of phaser synthesis is to realize a response as close as possible to the ideal response represented in Fig.~\ref{fig:phaser_eng}. Specifically, according to Sec.~\ref{sec:char_enhanc}, the phaser resolution, given by \eqref{eq:phaser_resol}, must be sufficiently high and its absolute bandwidth must be appropriate for the targeted application, and its transmission magnitude must be carefully equalized to combat frequency-dependent losses. The shape of the required phase response, represented by the Taylor series \eqref{eq:phase_Taylor} is most often quadratic, corresponding to a linear group delay, as required for instance in RTFT~(Sec.~\ref{sec:RTFT}), but it may also be cubic, quartic or of higher order, or more complex, as in the stepped group delay response used in the spectrum sniffer to be presented in Sec.~\ref{sec:ASP_apps}.

The synthesis of medium-type phasers is beyond the scope of this paper. We shall restrict here our attention to the case of network-type phasers. Little efforts have been dedicated to the synthesis such phasers to date, since filter synthesis systematically targets linear phase or zero-slope group delay responses, which exhibit no ASP capability. However, the important body of knowledge that has been built over the past decades in the synthesis of linear phase filters and all-pass equalizers may now be exploited and extended to the synthesis of actual phasers. The works reported in this area are too numerous to be exhaustively cited. They are described in many textbooks, such as for instance~\cite{Guillemin_SOB_1957,MYJ_AHP_1980,Cameron_Kudsia_Mansour_MFCS}. Some of the contributions most specifically related to phasers include the following. In 1963, Steenaart introduced the C-sections and D-sections, under the names of ``all-pass networks of the first order and second order'', respectively~\cite{Steenart_TMTT_01_1963}, as the first distributed implementations of the lumped-element all-pass lattice network~\cite{Cameron_Kudsia_Mansour_MFCS}. Cristal proposed a transmission-type cascaded coupled-line equalizer in~1966~\cite{Cristal_TMTT_06_1966} and a reflection-type equalizer using a circulator in~1969~\cite{Cristal_TMTT_01_1969}. Rhodes did extensive work on the synthesis of all-pass equalizers and linear-phase filters. For instance, in 1970, he proposed a method to synthesize linear-phase filters based on a recurrence formula generating Hurwitz polynomials with arbitrary phase versus frequency response~\cite{Rhodes_TMTT_06_1970}. Shortly later, he applied this method to synthesize a cross-coupled waveguide structure~\cite{Rhodes_TCT_03_1973}. However, this was before the advent of the coupling matrix technique~\cite{Cameron_Kudsia_Mansour_MFCS}, and the method led to high-order prototypes, due to the limitation of the synthesized transfer function. The phase polynomial generation procedure used in~\cite{Rhodes_TMTT_06_1970} was later refined by Henk~\cite{Henk_IJCTA_10_1981}, and seems to be the only one of the kind available in the literature. One may finally cite the recent work of Atia's group on the synthesis of reflection-type equalizers using coupled-resonators~\cite{Atia_TMTT_08_2002}. An excellent overview of all the developments on linear-phase filters and equalizers is available in~\cite{Cameron_Kudsia_Mansour_MFCS}.

We shall next discuss closed-form synthesis techniques that were recently developed for the three phasers selected in Sec.~\ref{sec:core_ASP_phaser}. These techniques are all based on the algorithm given in~\cite{Rhodes_TMTT_06_1970,Henk_IJCTA_10_1981}, and derived explicitly in~\cite{Gupta_IJCTA_0000}, for the least-mean-square approximate construction of a Hurwitz polynomial exhibiting a desired phase function. Assume a specified phase function $\phi(\Omega)$, where $\Omega$ will next represent the frequency variable in the lowpass domain. The aforementioned algorithm requires this phase function to be discretized in $N$~points, which leads to the frequency and phaser sets $\left\{\Omega_0,\Omega_1,\ldots\Omega_N\right\}$ and $\left\{\phi_0,\phi_1,\ldots\phi_N\right\}$, respectively. Based on these two sets, the algorithm, which reads
\begin{subequations}\label{eq:Henk_algo}
\begin{equation}\label{Eq:recurrence1}
\begin{split}
H_0(s)&=1\\
H_1(s)&=s+\alpha_0\\
&\vdots\\
H_N(s)&=\alpha_{N-1}H_{N-1}(s)
+\left(s^2+\Omega_{N-1}^2\right)H_{N-2}(s)\\
\end{split}
\end{equation}
with
\begin{equation}\label{Eq:recurrence2}
{\alpha _i} = \dfrac{{\Omega _{i + 1}^2 - \Omega _i^2}}{{{\alpha _{i - 1}} - \dfrac{{\Omega _{i + 1}^2 - \Omega _{i - 1}^2}}{{{\alpha _{i - 2}} - \dfrac{{\Omega _{i + 1}^2 - \Omega _{i - 2}^2}}{{\begin{array}{*{20}{r}}
   {}
   {\begin{array}{*{20}{r}}
    \ddots  & {}  \\
   {} & {{\alpha _0} - \Omega_{i+1}/\tan(\phi_{i+1}/2)}
\end{array}}
\end{array}}}}}}},\notag
\end{equation}
\begin{equation}
\alpha_0=\Omega_1/\tan(\phi_1/2),
\end{equation}
\end{subequations}
iteratively builds the $N^\text{th}$~degree Hurwitz polynomial $H_N(s)$ exhibiting the specified phase $\phi(\Omega)$, i.e. $\angle[H_N(j\Omega)]=\phi(\Omega)$. The order, $N$, of the polynomial will correspond to the order, and hence to the size, of the phaser structure, as will be shown next. Therefore, as the number of discretization points is proportional to the bandwidth over which the approximation~\eqref{eq:Henk_algo} must hold, the size of the phaser structure will tend to be proportional to its bandwidth.

\begin{figure*}[t!]
\center
\includegraphics[width=0.9\textwidth,page=9]{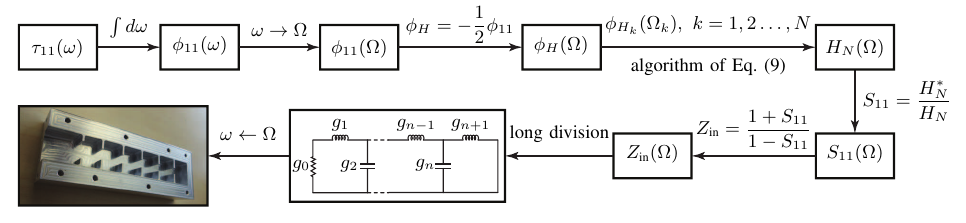}\\
\caption{Coupled-resonator (all-pass reflection-type) phaser synthesis procedure~\cite{Zhang_TMTT_08_2012}. The first row also applies to the the all-pass cascaded C-section (all-pass transmission-type) phaser design~\cite{Gupta_IJCTA_0000}, where the subscript pair~``11'' is then to be replaced by~``21''.}
\label{fig:allpass_synthesis_chart}
\end{figure*}

Let us first consider the synthesis of the all-pass reflection-type coupled-resonator phaser, shown in Fig.~\ref{fig:phasers}(c) and explained in Fig.~\ref{fig:network_phasers}(b). Its synthesis procedure was introduced in~\cite{Zhang_TMTT_08_2012} and is described by the flow chart of Fig.~\ref{fig:allpass_synthesis_chart}. Starting from a specified group delay function, $\tau_{11}(\omega)$, of the component's reflection function, $S_{11}(\omega)$, one first computes, analytically if possible or otherwise numerically, the corresponding phase function, $\phi_{11}(\omega)$, by integration. Next, using the band-pass~($\omega$) to low-pass~($\Omega$) mapping function corresponding to the phaser technology used, one obtains the phase function in the low-pass domain, $\phi_{11}(\Omega)$. Since the all-pass transfer function will read $S_{11}=H_N^*/H_N$, the phase of the sought Hurwitz polynomial is minus half that of $\phi_{11}(\Omega)$, i.e. $\phi_H(\Omega)=-\phi_{11}(\Omega)/2$. The function $\phi_H(\Omega)$ is then discretized over a reasonable (as small as possible) number $N$ of points, and the iterative algorithm of~\eqref{eq:Henk_algo} is applied to the resulting frequency and phase sets, yielding the polynomial \mbox{$H_N(s=j\Omega)$}, and hence $S_{11}(\Omega)$, from which the input impedance $Z_\text{in}(\Omega)=(1+S_{11})/(1-S_{11})$ follows. This impedance, expressed in terms of the ratio of two polynomials in~$\Omega$, is then written in the form of a long division, and mapped to the long division expression of the input impedance of the low-pass filter prototype, which provides the prototype's~$g_k$ parameters. From this point, conventional filter synthesis techniques~\cite{MYJ_AHP_1980} are applied to provide the final phaser structure.

The all-pass transmission-type cascaded C-section phaser, of the type shown in Fig.~\ref{fig:phasers}(g) and explained in Fig.~\ref{fig:network_phasers}(a) may be synthesized using the technique presented in~\cite{Gupta_IJCTA_0000}. The first part of this synthesis technique follows exactly the first row of the flow chart of Fig.~\ref{fig:allpass_synthesis_chart} after replacing the subscript pair~``11'' by~``21'', while its second part consists in determining the C-section lengths and coupling coefficients using formulas based on the analytical transfer functions of the C-section~(see synthesis flow chart of Fig.~7 in~\cite{Gupta_IJCTA_0000}). It should be noted that a synthesis technique for an alternative to this phaser, involving both \emph{commensurate} C-sections \emph{and} D-sections, was recently presented in~\cite{Zhang_IJRMCAE_TBP}. In this technique, the Hurwitz polynomial of the transfer function includes real roots corresponding to the C-sections and complex roots corresponding to the D-sections, and the C-section and D-section lengths and coupling coefficients are also determined from corresponding analytical transfer function.

\begin{figure}[h!]
\centering
\includegraphics[width=0.5\textwidth,page=10]{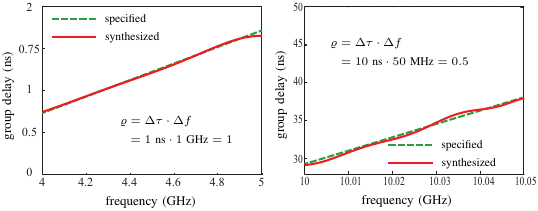}
\caption{Example of linear group delay all-pass phaser synthesis results. (a)~Cascaded C-section phaser of the type in Fig.~\ref{fig:phasers}(g) but with alternating C-sections to avoid inter-coupling~\cite{Gupta_IJCTA_0000} (synthesized: analytical, using a C-section based transfer function). (b)~Coupled-resonator phaser of the type in Fig.~\ref{fig:phasers}(c)~\cite{Zhang_TMTT_08_2012} (synthesized: full-wave).}
\label{fig:allpass_examples}
\end{figure}

Figures~\ref{fig:allpass_examples}(a) and (b) show the synthesized group delay responses for a cascaded C-section phaser and a coupled-resonator phaser, respectively. As announced in Sec.~\ref{sec:core_ASP_phaser}, the the latter allows operation over a much smaller bandwidth than the former, because of its reflection-type versus transmission-type nature.

\begin{figure*}[ht!]
\center
\includegraphics[width=0.9\textwidth,page=11]{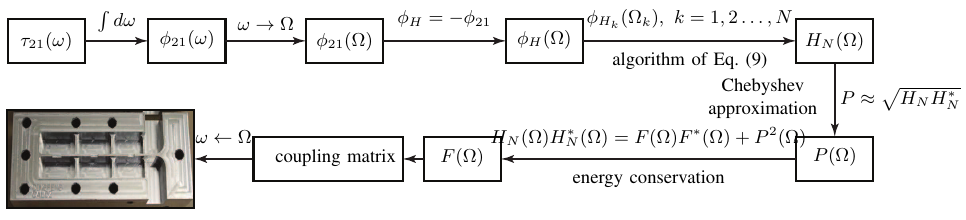}\\
\caption{Band-pass cross-coupled resonator phaser synthesis procedure.}
\label{fig:bandpass_synthesis_chart}
\end{figure*}

In the previous two all-pass type phasers, some magnitude imbalance is generally unavoidable and a resistive equalization network may have to be used to suppress it~(Sec.~\ref{sec:char_enhanc}). The band-pass transmission-type cross-coupled phaser, of the type shown in Fig.~\ref{fig:phasers}(i) and explained in Fig.~\ref{fig:network_phasers}(d), does not suffer from this drawback, because its phase and magnitude responses can be synthesized independently. Moreover, it may offer higher design flexibility, due to the absence (or minimality) of specification constraints in the stop-bands. Its synthesis procedure is described by the flow chart of Fig.~\ref{fig:bandpass_synthesis_chart}, where the transmission and reflection functions take the usual polynomial ratio forms
\begin{subequations}\label{eq:Sp_ccp}
\begin{equation}\label{eq:S21_ccp}
S_{21}(s)=\dfrac{P(s)}{H(s)},
\end{equation}
\begin{equation}\label{eq:S11_ccp}
S_{11}(s)=\dfrac{F(s)}{H(s)},
\end{equation}
\end{subequations}
\noindent where $H(s)$ is a Hurwitz polynomial. In these relations, the phase and magnitude responses will be exclusively controlled by~$H(s)$ and $P(s)$, respectively, so that we will have \mbox{$|H(s)|=1$} and $P(s)\in\Re$~\cite{Zhang_TMTT_03_2013}. Once the low-pass phase function, $\phi_{21}(\Omega)$, has been determined from the prescribed band-pass group delay function, $\tau_{21}(\omega)$, the required phase for $H(s=j\Omega)$ is found as $\phi_H(\Omega)=-\phi_{21}(\Omega)$, since $H(s)$ is in the denominator of $S_{21}$ in~\eqref{eq:S21_ccp}. The next step is to build the corresponding phase polynomial, $H(s)$, using the iterative algorithm~\eqref{eq:Henk_algo}, exactly as in the synthesis of the all-pass phasers. One may then specify the magnitude of $S_{21}(\Omega)$, $P(\Omega)$, in the phaser's pass-band, and this magnitude may approximate various functions, such as for instance that of a Chebyshev polynomial, \mbox{$P(\Omega)\approx\sqrt{H_N(\Omega)H_N^*(\Omega)}$}. Finally, the numerator of $S_{11}(\Omega)$, $F(\Omega)$ in~\eqref{eq:S11_ccp}, is determined by applying the energy conservation condition, from the initial assumption of a lossless system. From this point, the conventional coupling matrix technique can be applied and the corresponding phaser structure can be designed~\cite{Cameron_Kudsia_Mansour_MFCS}.

\begin{figure}[h!]
\centering
\includegraphics[width=0.5\textwidth,page=12]{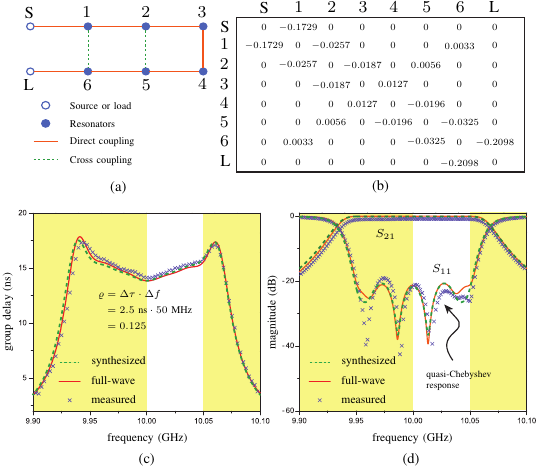}
\caption{Example of linear group delay band-pass phaser synthesis results, corresponding to the cross-coupled resonator structure of~Fig.~\ref{fig:phasers}(i). (a)~Topology. (b)~Coupling matrix. (c)~Up-chirp group delay response~(white range). (d)~Corresponding scattering parameters.}
\label{fig:bandpass_example}
\end{figure}

Figure~\ref{fig:bandpass_example} presents a cross-coupled phaser synthesis example. This design uses the simply folded topology shown in Fig.~\ref{fig:bandpass_example}(a) with the corresponding coupling matrix of Fig.~\ref{fig:bandpass_example}(b). The group delay response is plotted in Fig.~\ref{fig:bandpass_example}(c), while the magnitude response is shown in Fig.~\ref{fig:bandpass_example}(d), where the quasi-Chebyshev behavior is clearly apparent in $S_{11}$.

The following comparisons between the three considered phasers are useful, in connection with the discussions of~Sec.~\ref{sec:char_enhanc} on phaser characteristics. The cascaded C-section~[Fig.~\ref{fig:allpass_examples}(a)] has the highest resolution, $\varrho=1$, but the bandwidth that it requires to achieve a group delay swing in the order of~1~ns is in the order of~1~GHz. This may be appropriate if the application features a bandwidth of the same order, but if the application requires a resolution in the order of~10~MHz, then the relevant resolution collapses to~$0.01$, which is may be insufficient. In this case, the coupled-resonator~[Fig.~\ref{fig:allpass_examples}(a)] is clearly more indicated, with its resolution $\varrho=0.5$ for $50$~MHz or 0.1 for the relevant $10$~MHz bandwidth. The cross-coupled  resonator phaser~[Fig.~\ref{fig:bandpass_example}(c)], for the same absolute bandwidth ($\varrho=0.125$ over 50~GHz or 0.025 for 10~GHz), that is four times smaller, for the same number of resonators~(six). However, the topological degrees of freedom could be exploited here to increase the resolution, while avoiding a circulator.

In all the aforementioned synthesis techniques, iterative procedures may be applied to increase the synthesis bandwidth from the initial closed-form synthesis, as shown for instance in~\cite{Zhang_TMTT_08_2012}.

\section{ASP Applications}\label{sec:ASP_apps}

Although the area of microwave ASP is relatively recent, several ASP applications have already been reported in the past few years. We shall describe here three of them in some details and provide references to some others.

\begin{figure}[h!]
\centering
\includegraphics[width=\columnwidth,page=13]{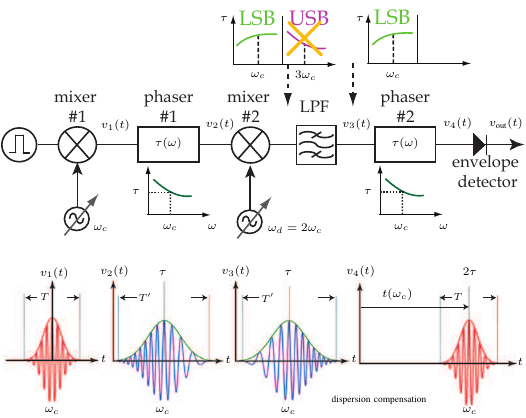}
\caption{Tunable pulse system, using mixer inversion for dispersion compensation~\cite{Abielmona_TMTT_11_2009}.}
\label{fig:tunable_pulse_delay_line}
\end{figure}

Let us first consider the \emph{tunable pulse system} that is depicted in Fig.~\ref{fig:tunable_pulse_delay_line} and that was proposed in~\cite{Abielmona_TMTT_11_2009} with a CRLH transmission line as a phaser. This system delays pulses by continuous and controllable amounts via a pair of synchronized local oscillators, one at~$\omega_c$ and the other one at $\omega_d=2\omega_c$, while avoiding pulse spreading using mixer inversion. The input signal is first mixed with the harmonic wave~$\omega_c$, which yields the modulated pulse~$v_1(t)$, with the same duration, $T$, as the input pulse. This pulse is then passed through the first phaser, whose center frequency coincides with $\omega_c$. This phaser does not require a linear group delay response; it may have any type of group delay response, such as for instance the CRLH-type one shown in the figure. The pulse at the output of the phaser, $v_2(t)$, has been delayed by an average amount  $\tau=\phi_1$~[Eq.~\eqref{eq:phi1}] (neglecting the delay induced by the mixer), chirped so as the exhibit the instantaneous frequency $\omega_2(t)$ and spread out in time to a duration $T'$ due to dispersion. Next, $v_2(t)$ is mixed with the harmonic wave $\omega_d=2\omega_c$, and passed through a band-pass filter which eliminates the upper side band (USB) to keep only the lower side band (LSB) of the mixer's output. The instantaneous frequency of the resulting pulse, $v_3(t)$, is $\omega_3(t)=2\omega_c-\omega_2(t)$, which is centered again at $\omega_c$, but whose chirp has now been \emph{inverted} [minus sign in front of $\omega_2(t)$]: this inversion process is called \emph{mixer chirp inversion}~\cite{Abielmona_TMTT_11_2009}. The pulse $v_3(t)$ is now pass through the second phaser, which must be \emph{identical} to the first one. Since the chirp of $v_3(t)$ has the opposite slope and an otherwise exactly identical shape as the response of the phaser, the phaser will exactly equalize (``un-chirp'') all the frequency components of $v_c(t)$, an effect called pulse compression in radar technology leading to an output signal $v_4(t)$ that has exactly retrieved the duration of the input pulse, $T$, but that has been delayed by an amount depending on $\omega_c$, based on the phaser's response, $t(\omega_c)$. Finally, the modulated pulse is envelop detected for based-band processing. Note that this pulse delay system does not suffer from any mismatch associated with tuning, like varactor controlled lines or components, since the control parameters, $\omega_c$ and $\omega_d$, are external to the transmission blocks.  This ASP tunable pulse delay system may be applied to ultra wideband processing, pulse position modulation~\cite{Nguyen_MWCL_08_2008} and compressive receivers~\cite{Abielmona_TMTT_11_2009}.

\begin{figure}
\includegraphics[width=\columnwidth,page=14]{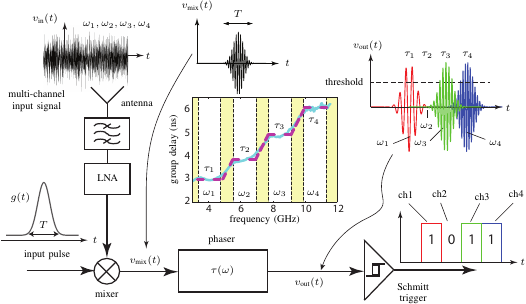}
\caption{Real-time spectrum sniffer, using a cascaded C-section stepped group delay phaser with the response shown at the center (dashed: specified, solid: experimental)~\cite{Nikfal_MWCL_11_2012}.} \label{Fig:sniffer}
\end{figure}

The next application example is the \emph{spectrum sniffer} that is described in Fig.~\ref{Fig:sniffer} and that was proposed in~\cite{Nikfal_MWCL_11_2012}. This system may be considered as a generalized frequency discriminator~[Fig.~\ref{fig:ASP_basic_effects}(b)]. Functionally, it ``listens'' to its radio environment through an omnidirectional antenna, and determines, in real time, the presence or absence of active channels in this environment so as to allow an associated communication system to opportunistically reconfigure itself to transmit in the available bands in a cognitive radio sense. The signal from the environment, $v_\text{in}(t)$, is multiplied with an auxiliary pulse, $g(t)$, after band-pass filtering and amplification, yielding $v_\text{mix}(t)$, which includes all the information on the environment's spectrum mixed up in time. The signal $v_\text{mix}(t)$ is then passed through a phaser with a \emph{stepped} group delay response, of the type already considered in Fig.~\ref{fig:ASP_basic_effects}(b), but including here four steps for four channels. In each frequency band, corresponding to an expected channel, centered at $\omega_k$~($k=1,2,3,4$), the group delay is flat, and hence the corresponding signal is neither distorted (i.e. could be demodulated) nor spread out in time, which will avoid channel overlap and subsequent interpretation errors at the output. The output signal, $v_\text{out}(t)$, is consists in a sequence of pulses that have been resolved in time, and that can next be passed through a Schmitt trigger to generate the binary information on the presence or absence of the channels. In the case of Fig.~\ref{Fig:sniffer}, channels~1, 3 and 4 are active, while channel~2 is not, at the observation time, and could therefore be temporarily used to maximize the data throughput. The authors are currently developing a system replacing the omnidirectional antenna by an electronically steered CRLH leaky-wave antenna~(LWA)~\cite{Abielmona_TAP_04_2011}, that is simultaneously temporally and spatially dispersive, as shall be seen in the next application example; this system will add the information of the angle-of-arrival of the different channels for even higher processing efficiency. It is worth noting that ASP, in the area of radiative systems, can also be used for RFID coding, where each tag consists in a different phaser with its specific group delay response~\cite{Gupta_AWPL_11_2011}.

\begin{figure}
\centering
\includegraphics[width=\columnwidth,page=15]{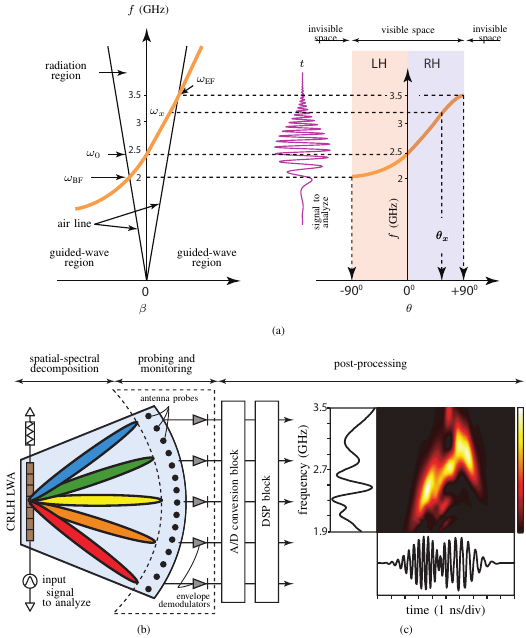}
\caption{Leaky-wave antenna (LWA) based real-time spectrum analyzer~\cite{Gupta_TMTT_04_2009}. (a)~Frequency-space mapping associated with temporal-spatial dispersion for the CRLH LWA employed in the RTSA. (b)~System schematic. (c)~Spectrogram example for a signal composed by the succession of two gaussian pulses with opposite chirps.}
\label{fig:RTSA}
\end{figure}

The last application example is the \emph{real-time spectrum analyzer~(RTSA)} that is described in Fig.~\ref{fig:RTSA} and that was proposed in~\cite{Gupta_TMTT_04_2009}. An RTSA is a system that determines, in real-time, the \emph{spectrogram} of signal, i.e. its \emph{joint time--frequency representation}, $X(\tau,\omega)=\mathcal{S}\left\{x(t)\right\}
=\int_{-\infty}^{+\infty}x(t)w(t-\tau)e^{-j\omega t}dt$, using a gating function~$w(t)$~\cite{Cohen_PIEEE_07_1989}. The spectrogram is most useful for \emph{nonstationary signals}, such as radar,
security, instrumentation, electromagnetic interference and compatibility and biological signals, where the sole frequency or time consideration of the signal may not provide a sufficiently informative representation of it. Here, the key component is a CRLH leaky-wave antenna~(LWA)~\cite{Caloz_Wiley_2006,Caloz_McrawHill_2011}, of the type shown in Fig.~\ref{fig:phasers}(f). As shown in Fig.~\ref{fig:RTSA}(a), such a LWA can be used to map the \emph{temporal frequencies} ($\omega$) of a broadband signal, whose spectrum lies within its radiation region, onto spatial frequencies ($\mathbf{k}$) or angles, $\theta$, via the antenna scanning law, $\theta=\sin^{-1}(c\beta/\omega)$, shown in the figure~\cite{Caloz_McrawHill_2011}; it is thus a simultaneously a temporally dispersive (phasing) and a spatially dispersive device. This double dispersive property is exploited in the RTSA as shown in Fig.~\ref{fig:RTSA}(b). The signal to analyze is injected into the CRLH LWA, which radiates its different time spectral components into different directions, following the scanning law of Fig.~\ref{fig:RTSA}(a). The corresponding waves are picked up with circularly arranged antenna probes, envelope detected, and processed in real time so as to build the spectrogram, shown in Fig.~\ref{fig:RTSA}(c) for the case of a signal composed by the succession of two gaussian pulses with opposite chirps. In contrast to digital spectrum analyzes, this RTSA only requires very light processing, features short acquisition times, can handle truly ultra wideband signals, and is easily scalable to the millimeter-wave frequency range.

The double dispersive and full-space properties of CRLH LWAs have recently led to a radiation pattern diversity multiple-input multiple-output (MIMO) communication system with drastically enhanced channel reliability and data throughput~\cite{MIMO_patent}. In this system, the LWAs are controlled by a processor, which monitors the received signal level in real-time and subsequently steers the LWAs so as achieve at all times the highest possible signal-to-noise ratio, and hence the highest possible channel capacity, in a real fluctuating wireless mobile environment.

\section{Challenges and Opportunities}

As pointed out in~Sec.~\ref{sec:intro_motiv}, ASP is not an established technology to systematically process microwave signals. This paper has shown that ASP technology features promising attributes and seems to possess a particularly great potential for millimeter-wave and terahertz applications, where conventional DSP-based systems are inefficient or even unapplicable. However, several challenges will have to be faced and various opportunities will have to be considered for this technology to succeed at a large scale.

Some of the most apparent challenges may be listed as follows: a)~realization of phaser characteristics for all the required applications (e.g. frequency resolutions finer than $100$~kHz, or fractional bandwidth broader than $150\%$); b)~closed-form synthesis of complex phasers (e.g. cross-coupled Cascaded C-section phasers); c)~fabrication of phasers in high millimeter-wave and terahertz frequencies, given that phasers are more sensitive to tolerances than filters, since they follow \emph{derivative} (phase derivative) as direct transfer function specifications; c)~development of optimal ASP modulation schemes for communications.

Some opportunities to meet these challenges and discover novel possibilities may include: i)~exploitation of other group delay functions (e.g. cubic, quartic or quintic, with various Taylor coefficients); ii)~utilization of novel, in particular nano-scale and multi-scale materials, for increased resolution; iii)~introduction of active elements to avoid restrictions related to the Foster reactance theorem~\cite{RWV_FWCE_1994} for higher dispersion diversity; iv)~introduction of nonlinearities, possibly using distributed semiconductor structures, for Kerr-type chirping, or self-phase modulation, in substitution to or in addition to dispersion, as in solitary waves~\cite{Agrawal_NLFO_1980}.

\section{Conclusion}

Based on the speculation that analog signal processing (ASP) may soon offer an alternative or a complement to dominantly digital radio schemes, especially at millimeter-wave and terahertz frequencies, we have presented this emerging area of microwaves in a general perspective. The preliminary research results are most promising. Dramatic progress has been made in the technology and synthesis of phasers, which are the key components in an ASP system. Several applications have already been reported and shown to offer distinct benefits over conventional DSP-based technology. It seems that microwave ASP has a great potential for future microwave, millimeter-wave and terahertz applications.

\section*{Acknowledgment}

This work was supported by NSERC Grant CRDPJ 402801-10 in partnership with Research In Motion (RIM).

\appendix[Impulse Response of a Linear-Chirp Phaser]
A phaser is considered here to be a linear system. The transfer function of a linear system is defined as~\cite{Papoulis_FIA}
\begin{equation}
H(\omega)
=\frac{\psi_\text{out}(\omega)}{\psi_\text{in}(\omega)},
\end{equation}
where $\psi_\text{in}(\omega)$ and $\psi_\text{out}(\omega)$ represent the input and output signals, respectively, in the spectral domain. The phaser transfer function, assuming unity magnitude, is obtained from the phase function~\eqref{eq:phase_Taylor} as
\begin{equation}\label{eq:transf_func}
H(\omega)=e^{j\phi(\omega)}
=e^{j\phi_0}e^{j\phi_1(\omega-\omega_0)}
e^{j\frac{\phi_2}{2}(\omega-\omega_0)^2}
e^{j\frac{\phi_3}{6}(\omega-\omega_0)^3}\ldots
\end{equation}
\indent A linear-chirp phaser is a phaser whose group delay (within a certain specified bandwidth) is a linear function of frequency or, equivalently according to~\eqref{eq:delay_Taylor}, whose phase is a quadratic function of frequency. Its transfer function~\eqref{eq:transf_func} includes then only the first three exponentials~(i.e.~$\phi_k=0$ for~$k>2$), where $\phi_2\neq 0$ to avoid the limit case of a non-dispersive (linear-phase) phaser:
\begin{equation}
H(\omega)
=e^{j\phi_0}e^{j\phi_1(\omega-\omega_0)}
e^{j\frac{\phi_2}{2}(\omega-\omega_0)^2}.
\end{equation}
\indent The corresponding impulse response, defined by~\cite{Papoulis_FIA},
\begin{equation}
h(t)=\int_{-\infty}^{+\infty}H(\omega)e^{j\omega t}d\omega,
\end{equation}
is calculated as
\begin{subequations}\label{eq:impulse_resp}
\begin{equation}
\begin{split}
h(t)
&=\int_{-\infty}^{+\infty}e^{j\phi_0}e^{j\phi_1(\omega-\omega_0)}
e^{j\frac{\phi_2}{2}(\omega-\omega_0)^2}
e^{j\omega t}d\omega\\
&=\gamma e^{j\frac{\beta t}{\phi_2}}e^{-j\frac{t^2}{2\phi_2}},
\end{split}
\end{equation}
with
\begin{equation}
\beta=\beta(\phi_2)
=\phi_1-\omega_0\phi_2,
\end{equation}
\begin{equation}
\gamma=\gamma(\phi_2)
=\sqrt{\dfrac{2\pi}{\phi_2}}
e^{j\left(\frac{\pi}{4}+\phi_0+\phi_1\omega_0
-\frac{\phi_1^2}{\phi_2}+\frac{\phi_2\omega_0^2}{2}
-\phi_2\omega_0^2\right)},
\end{equation}
\end{subequations}
where the tabulated result $\int_{-\infty}^{+\infty}e^{-(a\omega^2+b\omega)}d\omega=\sqrt{\pi/a}e^{\frac{b^2}{4a}}$ with \mbox{$a=-j\phi_2/2$} and $b=-j(t-\omega_0\phi_2)$ has been used.

\bibliographystyle{IEEEtran}
\bibliography{MM_RF_ASP_Caloz}


\end{document}